\documentclass{PoS}

\usepackage{latexsym}
\usepackage{graphicx}

\newcommand{\phib}{\ensuremath{\overline{\phi}}}
\newcommand{\etab}{\ensuremath{\overline{\eta}}}
\newcommand{\psib}{\ensuremath{\overline{\psi}}}
\newcommand{\DK}{Dirac-K\"{a}hler }
\newcommand{\beq}{\begin{equation}}
\newcommand{\eeq}{\end{equation}}

\PoS{PoS(LAT2005)006}

\title{Dirac-K\"{a}hler fermions and exact lattice supersymmetry}

\ShortTitle{Dirac-K\"{a}hler fermions and exact lattice supersymmetry}

\author{\speaker{Simon Catterall}\thanks{work
supported in part by DOE grant DE-FG02-85ER40237}\\
        Syracuse University\\
        E-mail: \email{smc@physics.syr.edu}}

\abstract{We discuss a new approach to putting supersymmetric theories on the
lattice. The basic idea is to start from a {\it twisted} formulation of
the underlying supersymmetric theory in which the fermions are represented
as grassmann valued antisymmetric tensor fields. The original supersymmetry
algebra is replaced by a twisted algebra which contains a scalar nilpotent
supercharge $Q$. Furthermore the action of the theory can then be written as
the $Q$-variation of some function. The case of ${\cal N}=2$ super Yang-Mills
theory in two dimensions is discussed in some detail. We then present our
proposal for discretizing this theory and derive the resultant
lattice action. The
latter is local, free of spectrum doubling, gauge invariant and 
preserves the scalar supercharge invariance
exactly. Some preliminary numerical results are then presented. The
approach can be naturally generalized to yield a lattice
action for ${\cal N}=4$ super Yang-Mills in four dimensions.
}

\FullConference{XXIIIrd International Symposium on Lattice Field Theory\\
		 25-30 July 2005\\
		 Trinity College, Dublin, Ireland}

\begin{document}

\section{Introduction}
Supersymmetric field theories exhibit many remarkable properties. Chief
among these are the cancellations which occur between boson and
fermion contributions in the perturbative calculation of physical
quantities. These cancellations eliminate many of the divergences
typical of quantum field theory and are at the heart of its use to
solve the gauge hierarchy problem \cite{wein}. Additionally, supersymmetric
versions of Yang Mills theory, while more tractable analytically than their
non-supersymmetric counterparts, exhibit many of the same features such as
confinement and chiral symmetry breaking \cite{prop}. Super Yang-Mills
theories with extended supersymmetry are also conjectured to be dual
to various supergravity theories in the limit of a large number of
colors and have been proposed as a regularization for M-theory \cite{dual,M}.

Since low energy physics is manifestly not  
supersymmetric it is necessary that this symmetry be broken at some  
energy scale. A set of non-renormalization theorems ensure that if SUSY is
not broken at tree level then it cannot be broken in any finite order of
perturbation theory see eg.~\cite{wein}. Thus the general expectation 
is that any breaking
should occur non-perturbatively. The lattice furnishes the only
tool for a systematic investigation of non-perturbative
effects in field theories and so significant effort has gone into
formulating SUSY theories on the lattice -- see \cite{general} and the
recent reviews \cite{feo},\cite{kaplan}.

Unfortunately, there  
are several barriers to such lattice formulations.
Firstly, supersymmetry is a spacetime symmetry which is generically  
broken by the discretization procedure. In this it resembles Poincare invariance
which is also not preserved in a lattice theory. However, unlike
Poincare invariance there is usually no SUSY analog of
the discrete translation and cubic rotation groups which are left unbroken on 
the lattice.
In the latter case the existence of these  
remaining discrete symmetries is sufficient to prohibit the appearance of   
relevant operators in the long wavelength lattice effective action which violate   
the {\it full} symmetry group. This ensures that Poincare
invariance is achieved automatically  
{\it without fine tuning} in the continuum limit.  
Since generic latticizations of supersymmetric theories do  
not have this property their effective actions typically  
contain relevant supersymmetry breaking interactions. 
To achieve a supersymmetric  
continuum limit then requires fine tuning the bare lattice couplings
of all these SUSY violating terms - typically a  
very difficult proposition.  
  
Secondly, supersymmetric theories necessarily involve fermionic fields which  
generically suffer from doubling problems when we attempt to define them on  
the lattice. The presence of extra fermionic modes furnishes yet
another source of supersymmetry breaking since typically they are not
paired with corresponding bosonic states. Furthermore,
most methods of eliminating the extra fermionic modes serve to  
break supersymmetry also.  
 
Recently there has been renewed interest in this problem stemming from
the realization that in certain classes of theory it may be possible to
preserve at least some of the supersymmetry {\it exactly} at finite
lattice spacing. It is hoped that this residual supersymmetry may
protect the lattice theory from the 
dangerous SUSY-violating radiative corrections that leading to fine
tuning.
In the case of
extended supersymmetry two approaches have been followed\footnote{
For examples of ${\cal N}=1$ models with exact SUSY see the recent
work \cite{feowz} and \cite{moto}}; in the first, pioneered
by Kaplan and collaborators, the lattice theory is
constructed by orbifolding a certain supersymmetric matrix model and
then extracting the lattice theory by expansion around some vacuum
state
\cite{kap,kap4,unsal,others}. 
The second approach relies on reformulating the supersymmetric theory in
terms of a new set of variables -- the {\it twisted} fields. In this
procedure a scalar nilpotent supercharge is exposed and it is the algebra
of this charge that one may hope to preserve under discretization
\cite{top},\cite{joel}, \cite{kaw2dtwist}.
This approach was initially used to construct 
and study lattice formulations of a variety of
low dimensional theories without gauge symmetry 
\cite{qm,qm_joel,wz2,wz2_joel,sigma}. Subsequently it was extended
to case of Yang-Mills theories by Sugino \cite{sugino1,sugino2}
although that construction was confined to low dimensional theories. 
In this
review we will confine ourselves to a discussion of a new
geometrical discretization of the twisted theory which 
accommodates both two and four dimensional super Yang Mills theories.

The basic idea in this approach is decompose the fermion fields as a set of
antisymmetric tensor fields which are subsequently interpreted as
components of single \DK field. This same procedure when applied to
the supercharges exposes a scalar nilpotent supercharge $Q$. Furthermore, the action of
the supersymmetric field theory can then be written as the $Q$-variation
of some function. Provided we can then maintain the nilpotent property
of $Q$ under discretization it is straightforward to write down a 
lattice action in terms of these twisted fields which is exactly
invariant under $Q$ at finite lattice spacing.
As a bonus we will find that that this twisted formulation of the
theory, being written entirely in geometrical terms, can be discretized
without encountering spectrum doubling. 

We first describe the nature of the twist in two dimensions and explain
how it is related to the use of \DK fermions. We then give a detailed
discussion of continuum ${\cal N}=2$ super Yang-Mills in two dimensions which
furnishes a simple example of the twisted approach. Our discretization
prescription is then introduced and a lattice action for this
theory constructed. The latter is local, free of spectrum doubling,
gauge invariant and preserves a single (twisted) supersymmetry at finite
lattice spacing. It is naturally formulated in terms of complex fields
and we discuss how this lattice model may be used to target the
true continuum theory. Preliminary results are then shown coming from
a RHMC simulation of this lattice model which indeed show an exact
$Q$-symmetry emerging at large coupling $\beta$. 
The technique can be generalized to provide a lattice action for
${\cal N}=4$ super Yang-Mills in four dimensions and this is then
discussed in some detail.

\section{The 2D Twist}
Consider a theory with ${\cal N}=2$ supersymmetry in two
dimensional Euclidean space. Such a theory contains
two Majorana supercharges $q^I_\alpha$ transforming under the global
symmetry group $SO(2)\times SO_R(2)$ where the subscript 
corresponds to two dimensional rotations and the superscript
describes the behavior under the R-symmetry corresponding to
rotating the two Majorana fields into one another.

The basic idea of twisting which goes back to
Witten \cite{tqft} is to introduce a new rotation group
\beq SO(2)^\prime={\rm diagonal\; subgroup}(SO(2)\times SO(2)_R)\eeq
and to decompose all fields now as representations of this new
rotation group. In simple terms what this means is that
whenever I do a rotation in the base space by some angle
I must do an equal rotation in the R-symmetry space. Thus it
is equivalent to treating the two indices $I$ and $\alpha$ as
equivalent. Thus the supercharges are to be interpreted as
{\it matrices} in this twisted picture \cite{tak,kato,kaw2dtwist,lat04}.
\beq q^{I}_\alpha\to q_{\alpha\beta}\eeq
It is now natural to expand such a matrix on a basis of products
of 2D gamma matrices
\beq q=QI+Q_\mu\gamma_\mu+Q_{12}\gamma_1\gamma_2\eeq
The fields $(Q,Q_\mu,Q_{12})$ are called the {\it twisted supercharges}.
The original SUSY algebra then implies a corresponding twisted algebra
which takes the form
\beq
\{q,q\}_{\alpha\beta}=4\gamma^\mu_{\alpha\beta}p^\mu
\eeq
In components this reads
\begin{eqnarray}
\{Q,Q\}&=&\{Q_{12},Q_{12}\}=\{Q,Q_{12}\}=\{Q_\mu,Q_\nu\}=0\\
\{Q,Q_\mu\}&=&P_\mu\\
\{Q_{12},Q_\mu\}&=&-\epsilon_{\mu\nu}P_\nu
\end{eqnarray}
Notice that the scalar component of the supercharge matrix
is nilpotent as previously advertised. It is also
important to realize that the twisted superalgebra implies that
the momentum $P$ is now the $Q$-variation of something -- it is
said to be $Q$-exact. This fact renders it plausible that the
entire energy momentum tensor may be $Q$-exact in such twisted
theories. Since $T_{\mu\nu}=\frac{\delta S}{\delta g_{\mu\nu}}$ this
would imply that the entire action of the theory could
be written in a $Q$-exact form $S=Q\Lambda$ \cite{lat04}. 
This turns out to be
true in essentially all the theories that have been studied.
Specifically we will show that the action of ${\cal N}=2$ super Yang-Mills
theory in two dimensions can be written in this way \cite{2dsuper}. 
Notice also that to match the four real supercharges of 
the original supersymmetric theory it is natural to take the
twisted supercharges to be real also. This in turn implies that
the supercharge matrix obeys a reality condition
\beq \overline{q}=q^\dagger=q^T\;{\rm in\;2D} \eeq

If the presence of
a nilpotent scalar fermionic charge and a corresponding $Q$-exact
action all sound reminiscent of BRST gauge fixing you would be right.
These twisted theories can be derived by gauge fixing an
underlying {\it topological} symmetry \cite{toprev}. 
If the gauge fixing function
is chosen carefully the resultant theory can be untwisted to
reveal an associated supersymmetric field theory. However, it must
be stressed that this correspondence is not one to one. In the
topological field theory the set of physical states must be
restricted in the usual way to those annihilated by the
scalar charge. In the language of the supersymmetric field theory
such a constraint would project one to the vacuum state. This
explains why such a theory contains no physically propagating degrees
of freedom -- it is a theory defined only on the moduli space of
classical solutions. Our proposal is not to impose this
physical state condition but merely to treat the twisting 
procedure as an exotic change of variables in the supersymmetric
field theory. Certainly in flat space the physical content
of the twisted and untwisted theories is the same.

Let us turn now to a discussion of the fields
in such a twisted theory. It should be clear
that if the supercharges can be written as matrix so can 
the fermions
\beq \Psi=\frac{\eta}{2}I+\psi_\mu\gamma_\mu+\chi_{12}\gamma_1\gamma_2\eeq
Thus the twisted theory will not contain spinors but antisymmetric
tensor fields. Indeed it is possible to abstract these p-form
fields and consider them as components of a geometrical object -- a
\DK field as first emphasized by Kawamoto and collaborators \cite{kaw2dtwist}.
\beq \Psi=(\frac{\eta}{2},\psi_\mu,\chi_{12})\eeq
It is a remarkable fact that
the original Dirac equation for 2 (Majorana) fermions is
then equivalent to the geometrical \DK equation \cite{banks}
\beq (d-d^\dagger)\Psi=0 \eeq
Here $d$ and $d^\dagger$ are the usual exterior derivative and
its adjoint. Their action on a generic \DK field $\omega$ is given
by 
\beq
d\omega = \left(0,\partial_\mu f, \partial_\mu f_\nu - \partial_\nu
f_\mu,\ldots\right)\eeq
and
\beq-d^\dagger\omega=\left(f^\nu,f^\nu_\mu,\ldots,0\right)_{;\nu}\eeq
The \DK equation can be derived from an action
\beq S_F=\left<\Psi^\dagger .(d-d^\dagger)\Psi\right> \eeq
where the dot product of two \DK fields $A$ and $B$ is defined by
\beq<A|B>=\int d^Dx\sqrt{g}\sum_p \frac{1}{p!}A^{\mu_1\ldots\mu_p}B_{\mu_1\ldots\mu_p}\eeq
Writing out these components explicitly 
and restricting ourselves to flat space we find 
\beq S_F=\frac{1}{2}\psi^\dagger_\mu \partial_\mu \eta+\frac{1}{2}\chi^\dagger_{\mu\nu}
\partial_{\left[\mu\right.}\psi_{\left.\nu\right]}\eeq
This equivalence of the Dirac equation to the \DK equation holds equally
when ordinary derivatives are replaced by gauge covariant derivatives.

The usual reason why the Dirac equation is preferred over the \DK
equation is that the latter naturally describes the propagation of
2 degenerate fermions in two dimensions. In a theory like QCD where
the bare quark masses are not usually thought of as equal this
appears problematic. However, a theory with ${\cal N}=2$ supersymmetry
in $D=2$ dimensions naturally requires such degenerate fermionic
states. Thus we have argued that for these special theories the
twisting procedure we have described is actually exactly equivalent
to the decomposition of the fermion fields in terms a single
\DK field. 
Furthermore the equivalence between the Dirac equation and the \DK
equation holds also in four dimensions where the latter equations
describes the propagation of four degenerate (Majorana) fermions.
Again, this is precisely the field content of ${\cal N}=4$ super
Yang-Mills theory and we should not be surprised then to find out
that the latter theory when suitably twisted can be expressed
in terms of the components of a single real \DK field
\cite{4dsuper},\cite{kaw4dtwist}.

Finally any theory invariant under the $Q$-symmetry
must contain superpartners for these fermionic fields.
In two dimensions we then expect to find the
following bosonic (commuting) fields
\beq (\phib,A_\mu,B_{12})\eeq
The nilpotency of $Q$ then requires that $Q^2=0$ up to gauge
transformations when acting on any of these component fields. If we
interpret $B_{12}$ as a multiplier field we see that in this way
we have generated the bosonic field content of 
${\cal N}=2$ SYM in 2D if we interpret each field as
taking values in the adjoint of some gauge group.
We turn now to an explicit construction of the twisted form
of this theory.
 
\section{Twisted ${\cal N}=2$ SYM in $D=2$}

In this section we render all these ideas concrete by writing down
the twisted formulation of ${\cal N}=2$ super Yang-Mills theory in
two (continuum) dimensions. The construction will expose the scalar
supersymmetric invariance explicitly and the $Q$-exact form of
the action. We will also show that the resulting on-shell action can be
rewritten in terms of the usual spinor degrees of freedom exhibiting
the fact that (in flat space) the twisted theory is nothing
but a change of variables in the original spinor based
super Yang-Mills model. The twisted gauge fermion is written
\cite{2dsuper},\cite{sugino1}
\begin{eqnarray}
S&=&\beta Q{\rm Tr}\int
d^2x\left(
\frac{1}{4}\eta[\phi,\phib]+2\chi_{12}F_{12}+
\right.\\
&+&\left.\chi_{12}B_{12}+\psi_\mu D_\mu \phib\right)\end{eqnarray}
where each field is in the adjoint of some gauge $U(N)$ group
$X=\sum_{a=1}^N X^a T^a$ and we employ antihermitian generators $T^a$.  
The action of $Q$ on the fields is given by
\begin{eqnarray}
QA_\mu&=&\psi_\mu\nonumber\\
Q\psi_\mu&=&-D_\mu\phi\nonumber\\
Q\phi&=&0\nonumber\\
Q\chi_{12}&=&B_{12}\nonumber\\
QB_{12}&=&[\phi,\chi_{12}]\nonumber\\
Q\phib&=&\eta\nonumber\\
Q\eta&=&[\phi,\phib]
\end{eqnarray}
Notice that $Q^2$ generates an (infinitessimal) gauge transformation
parametrized by the field $\phi$. Carrying out the $Q$-variation
and subsequently integrating out the multiplier field $B_{12}$ we find
the on-shell action
\begin{eqnarray}
S&=&\beta {\rm Tr}\int d^2x\left(
\frac{1}{4}[\phi,\phib]^2-\frac{1}{4}\eta [\phi,\eta]-F_{12}^2\right.\\
&-&D_\mu \phi D_\mu \phib-\chi_{12}[\phi,\chi_{12}]\\
&-&2\chi_{12}\left(D_1\psi_2-D_2\psi_1\right)-2\psi_\mu D_\mu\eta/2\\
&+&\left.\psi_\mu [\phib,\psi_\mu]\right)
\end{eqnarray}
The gauge and scalar part of this action are easily recognized to be
the usual bosonic sector of ${\cal N}=2$ super Yang-Mills. If I choose
the contour where $\phib^a=(\phi^a)^*$ it is clear that the bosonic action
is real and positive semi-definite. The fermionic sector looks a little
unconventional but by looking back at the previous discussion it can be
seen that it merely corresponds to a \DK decomposition of the
usual spinor action corresponding to two Majorana fermions. To convince
ourselves of the equivalence of this twisted formulation to the usual one
it is convenient to pick a Euclidean chiral 
basis for the two dimensional $\gamma$-matrices
\beq \begin{array}{cc}\gamma_1=
\left(\begin{array}{cc}
0&1\\
1&0
\end{array}\right)&
\gamma_2=\left(\begin{array}{cc}
0&i\\
-i&0
\end{array}\right)
\end{array}\eeq
and to subsequently define a Dirac spinor $\lambda$ as
\beq
\lambda=\left(\begin{array}{c}
\frac{\eta}{2}-i\chi_{12}\\
\psi_1-i\psi_2
\end{array}\right)\eeq
It is then a straightforward exercise to show that the Dirac
action $\lambda^\dagger \gamma.D\lambda$ reduces to the
kinetic terms in the twisted action we have previously uncovered.
Indeed the entire twisted action including the Yukawas can be
rewritten in terms of this Dirac spinor and can be
recognized as the usual fermionic sector of two
dimensional ${\cal N}=2$ super
Yang-Mills.
\beq S=\lambda^\dagger M(\phi) \lambda\eeq
where
\beq M=\gamma .D+\frac{\left(1+\gamma_5\right)}{2}[\phib,\cdots]
             -\frac{\left(1-\gamma_5\right)}{2}[\phi,\cdots]\eeq

\section{Discretization}
We will adopt this geometrical formulation as a useful starting off
point for constructing a lattice theory. It is convenient from the 
point of view of supersymmetry as the scalar component of the
twisted supersymmetry involves no derivatives and offers the
hope, therefore, of being compatible with a lattice structure.
Secondly, the use of \DK fermions is attractive as it is known
how to discretize the \DK action without inducing spectrum
doubling. Indeed, \DK fermions at the level of free field theory
are known to be equivalent to staggered fermions.  The usual
degeneracy of staggered fermions now becomes a plus -- it allows us to
replicated the degenerate flavors required by extended
supersymmetry. The use of \DK fermions to formulate (Hamiltonian) lattice
supersymmetric models was first proposed in \cite{schwim}.

Clearly, continuum scalar fields should map to fields defined on
sites in a (hypercubic) lattice, continuum vector fields to fields
defined on links and rank 2 antisymmetric fields to lattice
fields living on plaquettes. Notice that all lattice p-cubes whose
dimension is greater than zero come in two possible 
orientations corresponding to whether the defining vertices are written
down as an even or odd permutation of some standard ordering. In the
case of links this is just the usual statement that the link has two
possible directions. It is then natural (and as we will see later essential)
that this 2-fold orientation of the underlying p-cube can be
associated with the existence of two independent lattice fields for
each continuum field. These two fields can be represented by allowing
the lattice fields $X^a$ to be complex numbers thus promoting the
group from $U(N)$ to $GL(N,C)$.  
The lattice fields are also assigned gauge transformation properties
which generalize the usual transformation properties of site fields
and links in lattice gauge theory \cite{adjoint}:
\begin{eqnarray}
f(x)&\to& 
G(x)f_(x)G^{-1}(x)\\
f_\mu(x)&\to& G(x)
f_\mu(x)G^{-1}(x+\mu)\\
f_{12}(x)&\to&G(x)f_{12}(x)G^{-1}(x+1+2)
\end{eqnarray}
where $G=e^{\phi}$ is a lattice gauge transformation. Notice in the
naive continuum limit where $G\sim 1+\phi$ these reduce to the usual commutator
structure typical of adjoint fields. 
In general this leads to a point split or {\it shifted} commutator. For link
fields this looks like
\beq [\phi(x)f_\mu(x)-f_\mu(x)\phi(x+\mu)]\eeq
Notice that this commutator has a well-defined lattice gauge transformation
property.
Finally, we must replace the continuum gauge connection $A_\mu$ with
a Wilson gauge link $U_\mu(x)=e^{A_\mu}$. Notice now that
$A_\mu$ will be complex which implies that the gauge links
are {\it not} unitary at this stage in the construction.

Of course we have not yet finished. We need also to give a prescription
for replacing continuum covariant derivatives by difference operators.
Furthermore, as we will argue later, we are naturally led to
use covariant versions of forward and backward difference operators.
The results of these difference operators when acting on lattice
fields should also result in new lattice fields with well-defined
gauge transformation properties. Thus hitting a site field with
such a (forward) difference operator should lead to a field which gauge
transforms like a link field etc. It is not hard to show that
the following definition of
a forward difference operator acting on site and link
field satisfy these properties \cite{adjoint}
\begin{eqnarray}
D^+_\mu f(x)&=&U_\mu(x)f(x+\mu)-f(x)U_\mu(x)\\
D^+_\mu f_\nu(x)&=&U_\mu(x)f_\nu(x+\mu)-f_\nu(x)U_\mu(x+\nu)
\end{eqnarray}
They clearly reduce to continuum derivatives as $a\to 0$ and ensure
that the resulting fields gauge transform in the appropriate way.
Rather remarkably it is possible to write a covariant backward
difference which is truly adjoint to the above operator for
gauge invariant quantities. Notice that it will necessarily act
as a divergence operation (for a proof of this statement we refer
the reader to the paper by Aratyn et al \cite{adjoint})
\begin{eqnarray}
D^-_\mu
f_{\mu}(x)&=&f_\mu(x)U^\dagger_\mu(x)-U^\dagger_\mu(x-\mu)f_\mu(x-\mu)\\
D^-_\mu f_{\mu\nu}(x)&=&f_{\mu\nu}(x)U^\dagger_\mu(x+\nu)-U^\dagger_\mu(x-\mu)f_{\mu\nu}(x-\mu)
\end{eqnarray}
We have now almost all the ingredients needed
to construct our lattice super Yang-Mills
theory. The remaining key observation goes back to work by Becher, Joos,
Rabin and others \cite{becher,rabin} concerning discretizing actions written
in terms of p-forms and exterior derivatives. They give a topological
proof using results from homology theory that such actions
can be discretized in such a way that the resulting lattice theories
are completely free of spectrum doubling -- that is, the 
resulting lattice theories do
{\it not} contain any modes having no analog in the
continuum theory. However, such theories, like staggered fermion
models, will necessarily describe degenerate flavors of fermion. While this
is a problem for QCD it is a bonus in the supersymmetric theories
studied here as these degenerate flavors can represent the multiple fermions
inherent in extended supersymmetry. While these original papers
showed how to do this for ordinary derivatives, a generalization to
covariant derivatives was first given by Aratyn at al.\cite{adjoint} which
is the one employed here. The discretization prescription is given
explicitly by replacing the usual partial derivatives in the
continuum theory by appropriate difference operators:
\begin{center}
$D_\mu\to D^+$ if it acts like $d$\\
$D_\mu\to D^-$ if it acts like $d^\dagger$ 
\end{center}
Following this rule we find that the Yang-Mills field strength is 
discretized as follows 
\beq
F_{\mu\nu}(x)=D^+_\mu U_\nu(x)\to F^{\rm cont}_{\mu\nu}\;{\rm as}\;a\to 0\eeq
Notice it is automatically antisymmetric in its indices.
Using these ingredients we propose the following lattice action
for ${\cal N}=2$ super Yang-Mills
\begin{eqnarray}
S_L&=&\frac{1}{2}\beta Q{\rm Tr}\sum_{x}\left(
-\frac{1}{4}\eta^\dagger[\phi,\phib]-2\chi^\dagger_{12}F_{12}\right.\\
&-&\left.\chi^\dagger_{12}B_{12}-
\psi^\dagger_\mu D^+_\mu\phib+{\rm h.c}
\right)
\end{eqnarray}
Notice the necessity of introducing both fields and their hermitian
conjugates in order
to write down gauge invariant expressions. The corresponding lattice
supersymmetry is a simple generalization of the continuum ones \cite{2dsuper}
\begin{eqnarray}
QU_\mu&=&\psi_\mu\nonumber\\
Q\psi_\mu&=&-D^+_\mu\phi\nonumber\\
Q\phi&=&0\nonumber\\
Q\chi_{12}&=&B_{12}\nonumber\\
QB_{12}&=&[\phi,\chi_{12}]^{(12)}\nonumber\\
Q\phib&=&\eta\nonumber\\
Q\eta&=&[\phi,\phib]
\end{eqnarray}
where I have replaced $D_\mu\phi$ by $D^+_\mu\phi$ as required by
gauge invariance and the prime reflects the fact that we must
use a {\it shifted} commutator as previously discussed.
Carrying out the $Q$-variation and then integrating out the field
$B$ as in the continuum we obtain the on-shell lattice
action
\begin{eqnarray}
S_L&=&\frac{\beta}{2}{\rm Tr}\sum_x \left(
\frac{1}{4}[\phi,\phib]^2+F^\dagger_{12} F_{12}\right.\\
&-&\left.\frac{1}{4}\eta^\dagger[\phi,\eta]-\chi^\dagger_{12}[\phi,\chi_{12}]^{(12)}+
\psi^\dagger_\mu[\phib,\psi_\mu]^{(\mu)}\right.\\
&+&\left.(D^+_\mu \phi)^\dagger D^+_\mu \phib-
2\chi^\dagger_{12}\left(D^+_1\psi_2 -D^+_2\psi_1\right)\right.\\
&-&\left.2\psi^\dagger_\mu D^+_\mu\frac{\eta}{2}+{\rm h.c}\right)
\end{eqnarray}
This action is invariant under $Q$, finite gauge transformations and 
an additional $SO(1,1)$ symmetry
\beq \psi_\mu\to \lambda\psi_\mu,\;\;\eta ,\chi_{12}\to
\frac{1}{\lambda}\eta, \chi_{12}\eeq
\beq
\phib\to \frac{1}{\lambda^2}\phib,\;\;\phi\to \lambda^2\phi\eeq 
This symmetry is the analog of the usual exact $U(1)$ chiral symmetry of
staggered quarks. It is important to realize however that our coupling of
gauge fields to lattice p-form fields
does {\it not} correspond to the usual gauging of
staggered quarks.

It is illuminating to examine the gauge action in more detail. Substituting
the definitions of $F_{\mu\nu}$ we obtain
\beq
\beta {\rm Tr}\sum_x F^\dagger_{12}(x)F_{12}(x)\eeq
\beq
\beta{\rm Tr}\sum_x \left(2I-U_P-U^\dagger_P\right)+
\beta{\rm Tr}\sum_x \left(M_{12}+M_{21}-2I\right)
\eeq
where 
\beq
U_P=U_1(x)U_2(x+1)U^\dagger_1(x+2)U^\dagger_2(x)\eeq
resembles the usual Wilson plaquette term
and 
\beq M_{12}=U_1(x)U^\dagger_1(x)U_2^\dagger(x+1)U_2(x+1)\eeq
is a new zero area Wilson loop term which would vanish if the
link variables were restricted to unitary matrices. Notice the appearance of
the Wilson term depends crucially on the appearance of both
$F_{\mu\nu}$ and $F^\dagger_{\mu\nu}$ which lends some support to
the use of complex variables in the formulation of the theory.
Such a theory does not exhibit doubling of course and we might guess that
the exact supersymmetry will thus prevent doubles from occurring in
the fermionic sector also. We turn to this question now.

We can assemble the various p-form components of the \DK fermion
back into a 4 component object in the following way
\beq
\Psi=\left(\begin{array}{c}
\eta/2\\
\chi_{12}\\
\psi_1\\
\psi_2\end{array}\right) 
\eeq
The twisted fermion action can then be written in the matrix form
$\Psi^\dagger M\Psi$
where
\beq
M=\left(\begin{array}{cc}
-[\phi,]^{(p)}&K\\
-K^\dagger&[\phib,]^{(p)}
\end{array}\right)\eeq
and the operator
\beq
K=\left(\begin{array}{cc}
D^+_2&-D^+_1\\
-D^-_1&-D^-_2
\end{array}\right)\eeq
is just a discrete kinetic term for the fermions. The Yukawas appear
on the diagonal as indicated.
After integration we expect that the result will be a Pfaffian of
the operator $M$ (the reality conditions in the continuum are
important here).
To examine potential doubling problems in this
operator it suffices to examine $M$ in the free limit 
where its determinant reduces to the usual determinant of
a two dimensional {\it double free} scalar laplacian.
\beq {\rm Pf}(M)={\rm det}(K)={\rm det}(D^+_\mu D^-_\mu)\eeq
Thus, as advertised, the use of \DK fermions avoids the appearance of
spurious zeroes in the fermion operator.

\section{Continuum Limit}

The lattice formulation we have given requires a doubling of
degrees of freedom -- we have argued that this is quite natural
in any lattice theory and can be associated with the two possible
orientations of the underlying p-cube. We have parametrized this
doubling in terms of complex fields. However, the
target continuum theory that we are hoping to reproduce 
in the limit of vanishing lattice spacing corresponds to setting
\begin{eqnarray}
{\rm Im}X_\mu^a&=&0\;\; \mbox{all fields $X$ bar scalars}\\ 
\phib&=&-\phi^\dagger
\end{eqnarray}
At this point we must address how this continuum theory can be recovered
from our complexified theory. First, consider the effective
action that results from integrating out the grassmann variables.
It is clearly gauge invariant if we ignore possible phase
problems and replace the Pfaffian by a square root of
the (gauge invariant) determinant. It also clearly has the right
classical continuum limit. The only remaining question is whether
the Ward identities corresponding to the twisted
supersymmetry still hold (or more conservatively, hold in the
limit of vanishing lattice spacing with no additional fine
tuning) We conjecture that this may be so and have followed this
approach so far in our numerical work.

It is possible to make some progress in understanding why this might indeed be
true. First parametrize the general $GL(N,C)$ link field $U_\mu(x)$ in terms
of a unitary component $u_\mu(x)$ and a positive definite hermitian
piece $R_\mu(x)$ in the following way
\beq U_\mu(x)=R_\mu(x)u_\mu(x)\eeq
Now consider some Ward identity corresponding to the expectation
value of the $Q$-variation of
some operator $<QO>$. It is not hard to show that this object can be computed
{\it exactly} in the semi-classical limit $\beta\to\infty$. The argument
follows by recognizing that 
\beq \frac{\partial}{\partial\beta}<QO>=<Q(OS)>=0\eeq
where $Q$-exactness of the action has been used. Thus the expectation
value is independent of coupling $\beta$ and hence can be computed in
the semi-classical limit. This is a standard result of topological
field theory \cite{toprev}.
Now consider the second term in the gauge action
\beq \beta{\rm Tr}\sum_x \left(M_{12}+M_{21}-2I\right)\eeq
where
\beq M_{12}=U_1(x)U^\dagger_1(x)U_2^\dagger(x+1)U_2(x+1)\eeq
and insert the general decomposition of the gauge link given above. The result
for $M_{12}$ is
\beq M_{12}=R^2_1(x)R^2_2(x+1)\eeq
with a similar result for $M_{21}$. Notice it is independent
of the unitary piece $u_\mu(x)$. Thus in the limit $\beta\to\infty$
each $R_\mu(x)$ is driven to the identity and the Ward identity can be
computed in the space of purely unitary gauge links. Thus the Ward
identities of the truncated theory should hold at least for large
$\beta$ since they can be viewed as coming from the complexified theory
in the limit of infinite $\beta$. 
Of course large $\beta$ also corresponds to the limit of vanishing lattice
spacing and we see that this line of reasoning constitutes an
argument that the 
Ward identities in the truncated lattice theory should be realized
without fine tuning in the continuum limit. As we shall show in the next
section our preliminary numerical results are consistent with this.

\section{Simulations}

For our simulations we have taken the gauge links to be
unitary matrices taking values in $U(2)$. As in the continuum
theory the scalars $\phi$ and $\phib$ are taken to be
complex conjugates of each other\footnote{the antihermitian nature of our basis
$T^a$ actually ensures that $\phi^\dagger=-\phib$}. 
To regulate possible IR divergences associated with the
flat directions where $[\phi,\phib]=0$ we have
added a scalar mass term $m^2\phi^\dagger\phi$. This will
break supersymmetry softly. At the end of the calculation we should
like to send $m\to 0$ to recover the correct target theory. 
This bosonic action is real positive semi-definite, 
gauge invariant and clearly has the
correct naive continuum limit. To this gauge and scalar action
we should add the effective action gotten by integrating
over the grassmann valued fields. We will represent this as
\beq {\rm det}^{\frac{1}{4}} (M(U,\phi)^\dagger M(U,\phi)+m^2)\eeq
where $M$ is the lattice \DK operator introduced earlier.
Notice that we have added a gluino mass term for the fermions
which helps regulate potential zero modes. 
It is chosen to be identical in magnitude to the scalar mass. This ensures
the scalar and fermion determinants will cancel for zero gauge
coupling. In these preliminary simulations we have neglected the
issue of the phase of the fermion determinant which is reflected in
the appearance of just $M^\dagger M$ in the effective action. Finally,
the fourth root reflects the
Majorana nature of the continuum fermions. Notice that the determinant
is also gauge invariant.
  
To simulate this model we have used the RHMC algorithm
developed by Clark and Kennedy \cite{rhmc}.
The first step of this algorithm replaces the effective action
by an integration over auxiliary {\it commuting}
pseudofermion fields $F$, $F^\dagger$
in the following way
\beq {\rm det}^{\frac{1}{4}}(M^\dagger M+m^2)=\int DFDF^\dagger e^{-F^\dagger
\left(M^\dagger M+m^2\right)^{-\frac{1}{4}}F}\eeq
The key idea of RHMC is to use an optimal (in the minimax sense)
rational approximation to this inverse fractional power.
In practice it is important to use a partial fraction representation
of this rational approximation
\beq \frac{1}{x^{\frac{1}{4}}}\sim
\alpha_0+\sum_{i=1}^{N}\frac{\alpha_i}{x+\beta_i}\eeq
The numbers $\alpha_i$, $\beta_i$ for $i=1\ldots N$
can be computed offline using
the remez algorithm\footnote{many thanks to Mike Clark for providing us
with a copy of his remez code}. 
The resulting pseudofermion action becomes
\beq \alpha_0 F^\dagger F+\sum_{i=1}^NF^\dagger
\frac{\alpha_i}{M^\dagger M+m^2+\beta_i}F\eeq
It is thus just a sum of standard 2 flavor pseudofermion actions
with varying amplitudes and mass parameters. In principle
this pseudofermion action can now be used in a conventional
HMC algorithm to yield an {\it exact} simulation of the original
effective action \cite{hmc}. The final
trick needed to render this approach feasible is to utilize a {\it multi-mass
solver} to solve all $N$ sparse linear systems simultaneously and with
a computational cost determined primarily by the smallest shift 
$\beta_i$. We use a multi-mass version of the
usual conjugate gradient algorithm \cite{multi}. In practice for the
preliminary simulations shown here we have used $N=12$ and an approximation
that gives an absolute bound on the relative error of $10^{-6}$ over the
entire range of the spectrum (the lower eigenvalue is
given by $m^2$ and the upper can be determined from 
trial simulations using a Lanczos method). Further details will
be presented elsewhere \cite{soon}.  

The numerical results we present here come from
simulations where the lattice length takes the values
$L=2,4,8$ and the mass parameter $m=0.1,0.05$
over a range of coupling $\beta=0.5\to 4.0$. Notice that these
lattices are roughly equivalent to
$L=4,8,16$ for staggered quark simulations. We are currently extending
these calculations to larger lattices and smaller mass \cite{soon}.

We have devoted our initial runs to a check of the simplest
possible supersymmetric Ward identity -- namely the value of the
total action. If the latter corresponds to a $Q$-exact observable this
should vanish. This fact, together with the quadratic nature
of the fermion action 
allows us to compute the bosonic (gauge plus scalar) 
action exactly and for {\it all}
values of the coupling constant $\beta$ we find
\beq \beta<S_B>=\frac{3}{2}RL^2\eeq
where $R=4$ is the number of generators of $U(2)$. 
The following graph shows a plot of $\frac{\beta}{6L^2}<S_B>$ for a range of
couplings $\beta$ and three different lattice sizes using a mass $m=0.1$.
The bold lines shows the analytic prediction based on supersymmetry (for
clarity we have added multiples of $-0.25$ to the curves and lines to
split up the data from different lattice sizes)
While there are clearly deviations of order 4-5\% at small coupling these
appear to disappear at large $\beta$ in line with our
expectations. 

\begin{figure}[htb]
\begin{center}
\includegraphics[width=3.2in]
{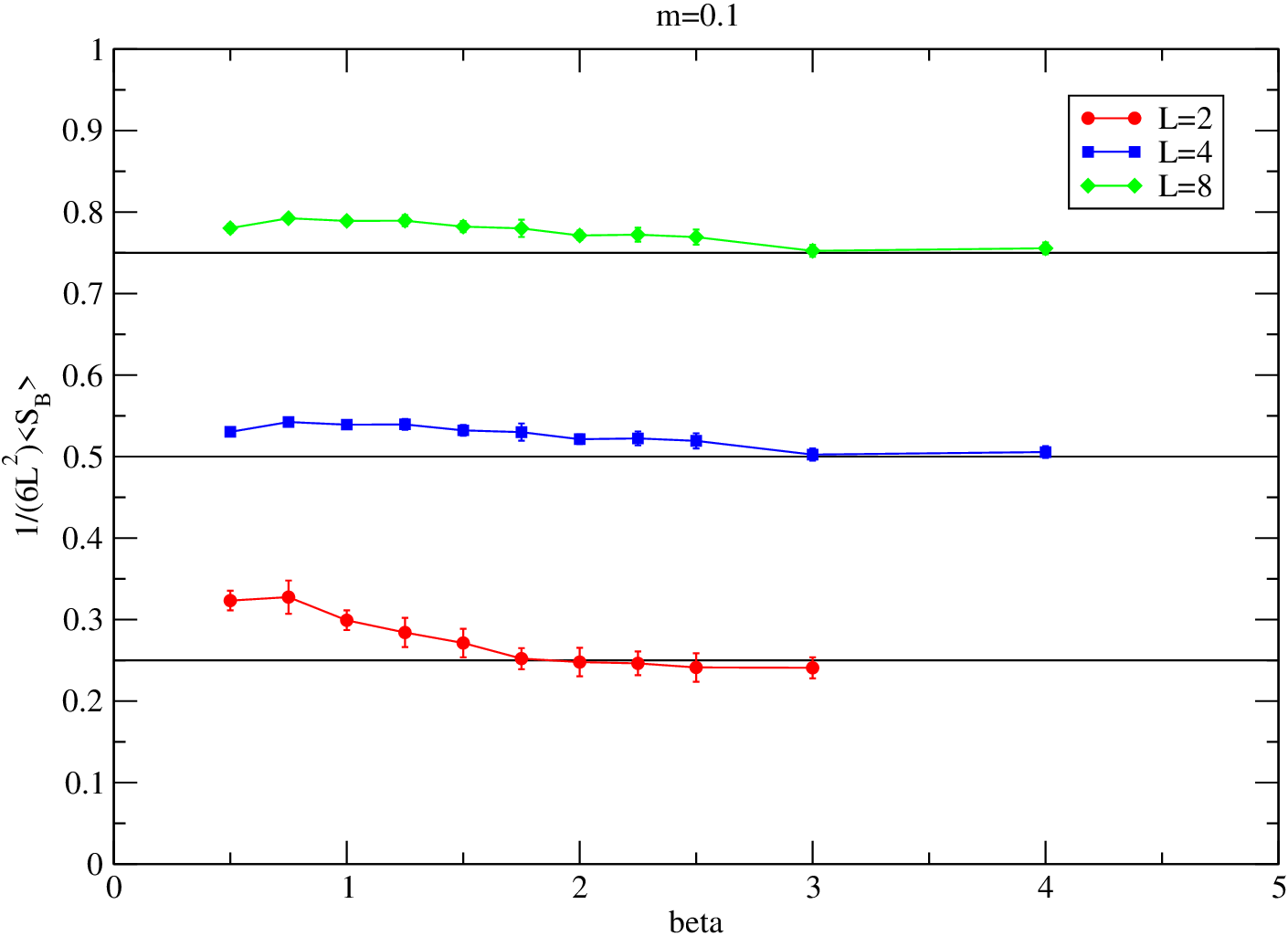}
\caption{$\frac{\beta}{6L^2}<S_B>$}
\end{center}
\end{figure}

We have also examined the gauge action and the string tension
determined by the smallest creutz ratio
$\sigma=-\ln{W(1,1)W(2,2)/(W(1,2)W(2,1))}$ and plot these as a function of
gauge coupling in the following plots ($m=0.1$ and $L=4$). 
These data are consistent with
the existence of a confining phase. 

\begin{figure}[htb]
\begin{center}
\includegraphics[width=3.2in]
{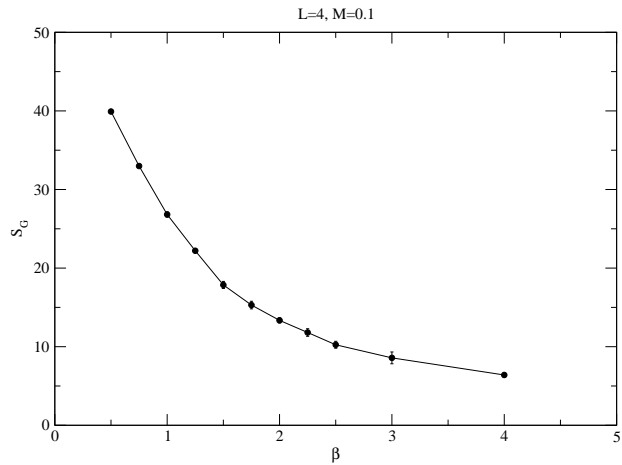}
\caption{Gauge action for $U(2)$ model} 
\end{center}
\end{figure}

\begin{figure}[htb]
\begin{center}
\includegraphics[width=3.2in]
{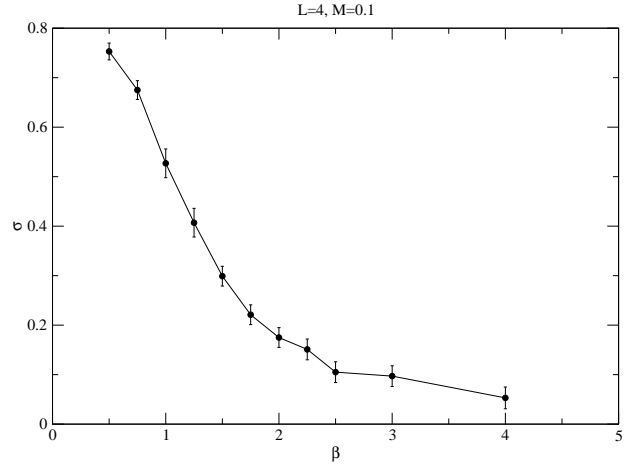}
\caption{String tension for $U(2)$ model}
\end{center}
\end{figure}

\section{Twisting ${\cal N}=4$ SYM in D=4}

The ${\cal N}=4$ super Yang-Mills theory in four
dimensions contains four Majorana spinors $\Psi^I_\alpha$
transforming under the Euclidean Lorentz group $SO(4)$ and
an additional $SO_R(4)$ R-symmetry group.
The K\"{a}hler-Dirac twist (see \cite{4dsuper},\cite{kaw4dtwist}) 
consists of decomposing those fields into representations of
the diagonal subgroup
\beq SO(4)^\prime={\rm diag}(SO(4)\times SO(4)_R)\eeq
As for two dimensions this twisting procedure is
equivalent to regarding the supercharges and fermions as matrices:
\begin{eqnarray}
\Psi&=&\eta
I+\psi_\mu\gamma_\mu+\frac{1}{2!}\chi_{\mu\nu}\gamma_\mu\gamma_\nu+\\
&+&\frac{1}{3!}\theta_{\mu\nu\lambda}\gamma_\mu\gamma_\nu\gamma_\lambda
+\frac{1}{4!}\kappa_{\mu\nu\lambda\rho}\gamma_\mu\gamma_\nu\gamma_\lambda\gamma_\rho
\end{eqnarray}
To match the 16 real supercharges of the
continuum theory it is natural to take these
antisymmetric tensor fields real. They can then be regarded as
the 16 grassmann components of a single real K\"{a}hler-Dirac field just
as for two dimensions. There will now be a scalar nilpotent
supercharge\footnote{the full twisted superalgebra was given by
D'Adda et al in\cite{kaw4dtwist}} and corresponding bosonic
superpartners 
$(\phib,A_\mu,B_{\mu\nu},W_{\mu\nu\lambda},C_{\mu\nu\lambda\rho})$ 
In the end $B$ and $C$ will turn out to be multiplier fields and are
integrated out of the final theory leaving two scalars $\phib$, $\phi$
(which arises as a gauge parameter as in two dimensions), one vector
$A_mu$ and the 4 independent components of $W_{\mu\nu\lambda}$.
These latter four fields will combine with the two scalars to yield the
usual 6 scalars of ${\cal N}=4$ super Yang-Mills. 

The form of the $Q$-variations of these fields under the
scalar supersymmetry can be obtained in analogy to the situation
in two dimensions.
\begin{eqnarray}
Q\phib&=&\eta\;\;Q\eta=[\phi,\phib]\nonumber\\
QA_\mu&=&\psi_\mu\;\;Q\psi_\mu=-D_\mu\phi\nonumber\\
QB_{\mu\nu}&=&[\phi,\chi_{\mu\nu}]\;\;Q\chi_{\mu\nu}=B_{\mu\nu}\nonumber\\
QW_{\mu\nu\lambda}&=&\theta_{\mu\nu\lambda}\;\;Q\theta_{\mu\nu\lambda}=
[\phi,W_{\mu\nu\lambda}]\nonumber\\
QC_{\mu\nu\lambda\rho}&=&[\phi,\kappa_{\mu\nu\lambda\rho}]\;\;Q\kappa_{\mu\nu\lambda\rho}=
C_{\mu\nu\lambda\rho}\nonumber\\
Q\phi&=&0
\label{Q}
\end{eqnarray}
In deciding how to write down these transformations one is guided by
the identification of physical fields. If a field is ultimately to
be identified as a propagating field it is necessary to make its
$Q$-variation yield another field. Multiplier fields vary into
commutators. Notice again that the square of the supercharge is 
an infinitessimal gauge transformation parametrized by the field $\phi$.

We hypothesize that, following on from our general arguments, it should
be possible to write the action as $Q$-exact function
\beq S=\beta Q\Lambda\eeq
where the
correct gauge fermion of ${\cal N}=4$ super Yang-Mills turns out
to be
\begin{eqnarray}
\Lambda&=&\int d^4x {\rm Tr}\left[
\chi_{\mu\nu}\left(F_{\mu\nu}+\frac{1}{2}B_{\mu\nu}-
\frac{1}{2}[W_{\mu\lambda\rho},W_{\nu\lambda\rho}]\right.\right.\nonumber\\
&+&\left.\left.D_\lambda W_{\lambda\mu\nu}\right)\right.\nonumber\\
&+&\left.\psi_\mu D_\mu\phib+\frac{1}{4}\eta[\phi,\phib]+
\frac{1}{3!}\theta_{\mu\nu\lambda}[W_{\mu\nu\lambda},\phib]\right.\nonumber\\
&+&\left.\frac{1}{4!}\kappa_{\mu\nu\lambda\rho}\left(
\sqrt{2}D_{\left[\mu\right.}W_{\left.\nu\lambda\rho\right]}+
\frac{1}{2}C_{\mu\nu\lambda\rho}\right)
\right]
\label{qact}
\end{eqnarray}\\
Several of these terms are in common with the gauge fermion of ${\cal N}=2$ super
Yang-Mills theory in two dimensions. The new ones involve the $3$ and $4$-form
fields. Of these the terms involving derivatives must be present to generate the
correct \DK action for the twisted fermions (and will simultaneously
generate the appropriate kinetic terms for
the W-field).
The commutator term involving $W_{\mu\nu\lambda}$ coupled to
$\chi_{\mu\nu}$ will generate a quartic potential for the W-field analogous to
that generated for the scalars $\phi$ and $\phib$. This will allow contact
to be made eventually with the supersymmetric theory
where one expects the scalars and W-field to play similar roles. 
In the same way the
commutator term involving $\phib$ and $W$ will also generate the
necessary mixed quartic
couplings between the scalars and the W-field.
Carrying out the $Q$-variation leads to the following action
\beq S=\beta\left(S_B+S_F+S_Y\right)\eeq
where the piece of the action $S_B$ involving the bosonic fields takes the form
\begin{eqnarray}
S_B&=&\int d^4x {\rm Tr}\left[
B_{\mu\nu}\left(F_{\mu\nu}-
\frac{1}{2}[W_{\mu\lambda\rho},W_{\nu\lambda\rho}]+
D_\lambda W_{\lambda\mu\nu}+
\frac{1}{2}B_{\mu\nu}\right)\right.\nonumber\\
&-&\left.D_\mu\phi D_\mu\phib+\frac{1}{4}[\phi,\phib]^2-
\frac{1}{3!}[\phi,W_{\mu\nu\lambda}][\phib,W_{\mu\nu\lambda}]\right.\nonumber\\
&+&\left.\frac{1}{4!}C_{\mu\nu\lambda\rho}\left(
\sqrt{2}D_{\left[\mu\right.}W_{\left.\nu\lambda\rho\right]}+
\frac{1}{2}C_{\mu\nu\lambda\rho}\right)
\right]
\end{eqnarray}
and the fermion kinetic terms are given by $S_F$ with
\begin{eqnarray}
S_F&=&\int d^4 x {\rm Tr}\left[
-\chi_{\mu\nu}D_{\left[\mu\right.}\psi_{\left.\nu\right]}
-\chi_{\mu\nu}D_\lambda\theta_{\lambda\mu\nu}
-\eta D_\mu\psi_\mu-
\frac{\sqrt{2}}{4!}\kappa_{\mu\nu\lambda\rho}D_{\left[\mu\right.}
\theta_{\left.\nu\lambda\rho\right]}
\right]
\end{eqnarray}
and $S_Y$ contains the Yukawa couplings
\begin{eqnarray}
S_Y&=&\int d^4x {\rm Tr}\left[
-\frac{1}{4}\eta[\phi,\eta]
-\frac{1}{2}\frac{1}{4!}
\kappa_{\mu\nu\lambda\rho}[\phi,\kappa_{\mu\nu\lambda\rho}]-
\frac{1}{2}\chi_{\mu\nu}[\phi,\chi_{\mu\nu}]\right.\nonumber\\
&+&\left.\psi_\mu[\phib,\psi_\mu]+
\frac{1}{3!}\theta_{\mu\nu\lambda}[\phib,\theta_{\mu\nu\lambda}]
\right.\nonumber\\
&+&\left.\frac{1}{3!}\eta[\theta_{\mu\nu\lambda},W_{\mu\nu\lambda}]-
\frac{\sqrt{2}}{4!}\kappa_{\mu\nu\lambda\rho}
[\psi_{\left[\mu\right.},W_{\left.\nu\lambda\rho\right]}]\right.\nonumber\\
&+&\left.\chi_{\mu\nu}[\theta_{\mu\lambda\rho},W_{\nu\lambda\rho}]-
\chi_{\mu\nu}[\psi_\lambda,W_{\lambda\mu\nu}]
\right]
\end{eqnarray}
Integrating over the multiplier fields $B_{\mu\nu}$ and $C_{\mu\nu\lambda\rho}$
and subsequently 
utilizing the Bianchi identity leads to a new bosonic action
of the form
\begin{eqnarray}
S_B&=&\int d^4 x {\rm Tr}\left[
-\frac{1}{2}\left(
\left(F_{\mu\nu}-
\frac{1}{2}[W_{\mu\lambda\rho},W_{\nu\lambda\rho}]\right)^2
+\left(D_\lambda W_{\lambda\mu\nu}\right)^2+
\frac{2}{4!}\left(D_{\left[\mu\right.}
W_{\left.\nu\lambda\rho\right]}\right)^2\right)
\right.\nonumber\\
&-&\left.D_\mu\phi D_\mu\phib+\frac{1}{4}[\phi,\phib]^2-
\frac{1}{3!}[\phi,W_{\mu\nu\lambda}][\phib,W_{\mu\nu\lambda}]
\right]
\end{eqnarray}
At this point it is useful to trade the $W$, $\theta$, and $\kappa$ fields
for new variables which will allow contact to be made between this
theory and one of the conventional twists of ${\cal N}=4$ super Yang-Mills.
We write
\begin{eqnarray}
W_{\mu\nu\lambda}&=&\epsilon_{\mu\nu\lambda\rho}V_\rho\nonumber\\
\theta_{\mu\nu\lambda}&=&\epsilon_{\mu\nu\lambda\rho}\psib_\rho\nonumber\\
\kappa_{\mu\nu\lambda\rho}&=&\epsilon_{\mu\nu\lambda\rho}\etab
\end{eqnarray}
In terms of these variables and after an additional integration
by parts the bosonic action reads (notice that terms linear in $F_{\mu\nu}$
cancel)
\begin{eqnarray}
S_B&=&\int d^4x {\rm Tr}\left[
-\frac{1}{2}F^2_{\mu\nu}-\frac{1}{2}[V_\mu,V_\nu]^2
-\left(D_\mu V_\nu\right)^2\right.\nonumber\\
&-&\left.D_\mu\phi D_\mu\phib+\frac{1}{4}[\phi,\phib]^2-
[\phi,V_{\mu}][\phib,V_{\mu}]
\right]
\end{eqnarray}
Making the additional rescalings $\chi\to 2\chi$ and $\etab\to \frac{1}{\sqrt{2}}\etab$ we find the fermion kinetic term takes the form
\begin{equation}
S_F=\int d^4 x {\rm Tr}\left[
-2\chi_{\mu\nu}D_{\left[\mu\right.}\psi_{\left.\nu\right]}
-2\chi^*_{\mu\nu}D_{\left[\mu\right.}\psib_{\left.\nu\right]}
-2\frac{\eta}{2} D_\mu\psi_\mu
-2\frac{\etab}{2} D_\mu\psib_\mu
\right]
\end{equation}
where 
$\chi^*_{\mu\nu}=\frac{1}{2}\epsilon_{\mu\nu\lambda\rho}\chi_{\lambda\rho}$
is the dual field.
In these variables the Yukawa's take on the more symmetrical
form
\begin{eqnarray}
S_Y&=&\int d^4x {\rm Tr}\left[
-\frac{\eta}{2}[\phi,\frac{\eta}{2}]
-\frac{\etab}{2}[\phi,\frac{\etab}{2}]
-2\chi_{\mu\nu}[\phi,\chi_{\mu\nu}]\right.\nonumber\\
&+&\left.\psi_\mu[\phib,\psi_\mu]+
\psib_\mu[\phib,\psib_\mu]
\right.\nonumber\\
&+&\left.2\frac{\eta}{2}[\psib_{\mu},V_\mu]-2\frac{\etab}{2}[\psi_\mu,V_\mu]
\right.\nonumber\\
&+&4\left.\chi_{\mu\nu}[\psib_{\mu},V_{\nu}]-
4\chi^*_{\mu\nu}[\psi_\mu,V_{\nu}]
\right]
\end{eqnarray}
The new action can be recognized as the twist of ${\cal N}=4$ super
Yang-Mills due to Marcus \cite{marcus}. 

Finally we will show how this twisted model may be reinterpreted in
terms of the usual formulation of ${\cal N}=4$ super Yang-Mills theory.
First, concentrate on the bosonic action and introduce the new
fields ($\phi=\phi_1+i\phi_2$)
\begin{eqnarray}
X^\mu&=&V_\mu\;\;\mu=0\ldots 3\nonumber\\
X^4&=&\phi_1\nonumber\\
X^5&=&\phi_2
\end{eqnarray}
Then the bosonic action may be trivially rewritten
as
\beq
S_B=-\frac{1}{2}F^2_{\mu\nu}-\left(D_\mu X^i\right)^2-
\frac{1}{2}\sum_{ij}[X_i,X_j]^2
\eeq
Notice this is real, positive semidefinite on account of the
antihermitian basis for the fields. It is also precisely the bosonic
sector of the ${\cal N}=4$ super Yang-Mills action with $X^i$
the usual $6$ real scalars of that theory \cite{4dspinor}. 

Next let us turn our attention to the fermion kinetic term.
From our previous discussion 
it should be clear that the twisted fermion kinetic term is nothing
more than the component expansion of a \DK action:
\beq S_F=\frac{1}{2}{\rm Tr}\Psi^\dagger \gamma .D \Psi \eeq
where $\Psi$ is the \DK field defined earlier with $\eta\to\eta/2$, 
$\theta_{\mu\nu\lambda}\to\epsilon_{\mu\nu\lambda\rho}\psib_\rho$
and $\kappa_{\mu\nu\lambda\rho}\to\epsilon_{\mu\nu\lambda\rho}\etab/2$.
Naively such an action appears to describe a theory with {\it four}
Dirac spinor fields - rather than the two one would have expected
for ${\cal N}=4$ super Yang-Mills. However it is evident that $\Psi$
obeys a reality condition if its associated \DK field is real. This
reduces the action to that of two degenerate
Dirac fermions. To see this in detail let us 
adopt a Euclidean chiral basis for the $\gamma$-matrices
\beq
\begin{array}{cc}
\gamma_0=\left(\begin{array}{cc}
0&I\\
I&0\end{array}\right)
&
\gamma_i=\left(\begin{array}{cc}
0&-i\sigma_i\\
i\sigma_i&0\end{array}\right)
\end{array}
\eeq
It is straightforward to see that the $\gamma$ matrices (and any
products of them) obey a reality condition
\beq
\gamma_\mu^*=C\gamma_\mu C^{-1}\eeq
where the matrix $C$ is given by
\beq C=\left(\begin{array}{cc}
\sigma_2&0\\
0&\sigma_2\end{array}\right)\eeq
With real p-form coefficients this implies
a reality condition on $\Psi$ itself
\beq \Psi^*=C\Psi C^{-1}\eeq
This in turn implies that $\Psi^\dagger=C\Psi^T C^{-1}$ which makes it
clear that the result of integrating over the
\DK field $\Psi$ should be the Pfaffian of the \DK operator
(let us neglect the Yukawa couplings for the moment).
This, in turn, will correspond to the product of {\it two} 
Dirac determinants.

To see this in more detail one can use the
reality condition to show that successive columns $\Psi^{(n)}$ of $\Psi$ 
are not independent but are charge conjugates of each other
\beq \Psi^{(2)}=C\left(\Psi^{(1)}\right)^*\;\;
\Psi^{(4)}=C\left(\Psi^{(3)}\right)^*\eeq
These conditions allow us to rewrite
the twisted fermion kinetic term in the conventional form
\beq \frac{1}{2}\sum_{\alpha=1,2}\lambda^{\dagger}_\alpha \gamma .D \lambda_\alpha\eeq
where the spinors are read off as the first and third columns of
the $\Psi$ matrix in this chiral basis:
\beq
\begin{array}{cc}
\lambda_1=\left(\begin{array}{c}
\frac{\eta}{2}-\frac{\etab}{2}+2i\chi^+_{03}\\
-2\chi^+_{02}+2i\chi^+_{01}\\
\left(\psi_0+\psib_0\right)+i\left(\psi_3+\psib_3\right)\\
-\left(\psi_2+\psib_2\right)+i\left(\psi_1+\psib_1\right)
\end{array}\right)
&
\lambda_2=\left(\begin{array}{c}
\left(\psi_0-\psib_0\right)+i\left(-\psi_3+\psib_3\right)\\
\left(\psi_2-\psib_2\right)+i\left(\psib_1-\psi_1\right)\\
\frac{\eta}{2}+\frac{\etab}{2}-2i\chi^-_{03}\\
2\chi^-_{02}-2i\chi^-_{01}
\end{array}\right)\\
\end{array}
\eeq
Here, $\chi^{\pm}=\frac{1}{2}\left(\chi\pm\chi^*\right)$ are the usual
self-dual and antiself-dual parts of the field. 
The Yukawa's can also be put in the general form
\beq \frac{1}{2}\sum_{\alpha=1,2}\lambda^\dagger_\alpha
C^\alpha_i\Gamma^i[X^i,\lambda_\alpha]\eeq
where $\Gamma^i=\left\{I,\gamma_5,\gamma_\mu\gamma_5,\mu=0\ldots 4\right\}$
and the coefficients $C^\alpha_i$ are just numbers. This is just
the structure expected of ${\cal N}=4$ super Yang-Mills.
Notice that both Dirac operators $M^\alpha$ including the Yukawa
interactions possess the symmetry
\beq \left(D^\alpha\right)^*=C D^\alpha C^{-1}\eeq
which shows that their eigenvalues come in complex conjugate
pairs and hence the associated determinants are positive definite.

\section{Lattice ${\cal N}=4$ super Yang-Mills}
The lattice action is obtained by discretization of the 
$Q$-exact action given in eqn.~\ref{qact}. Notice it is
important {\it not} to use the same dual variables as Marcus 
in discretizing the theory as
the duality transformation is {\it not} compatible with the lattice
gauge transformations we have described. With
that proviso the prescription is essentially the
same as for the two dimensional theory \cite{4dsuper}. We list it now
for a $D$-dimensional theory.
\begin{itemize}
\item A continuum p-form field $f_{\mu_1\ldots \mu_p}(x)$ will be mapped
to a corresponding lattice field associated with
the $p$-dimensional hypercube at lattice site $x$ 
spanned by the (positively directed) unit vectors $\{\mu_1\ldots\mu_p\}$. 
\item Such a lattice
field is taken to transform under gauge transformations in the following way
\beq
f_{\mu_1\ldots\mu_p}(x)\to G(x)f_{\mu_1\ldots \mu_p}(x)G^{-1}(x+e_{\mu_1\ldots\mu_p})\eeq
where the vector $e_{\mu_1\ldots\mu_p}=\sum_{j=1}^p\mu_j$. 
\item To construct gauge invariant quantities we will need to introduce both
$f_{\mu_1\ldots\mu_p}$ 
and its hermitian conjugate $f^\dagger_{\mu_1\ldots\mu_p}(x)$. The
latter transforms as
\beq
f^\dagger_{\mu_1\ldots\mu_p}(x)\to G(x+e_{\mu_1\ldots\mu_p})
f_{\mu_1\ldots \mu_p}(x)G^{-1}(x)\eeq
Since for all fields bar the gauge field we will assume
$f_{\mu_1\ldots\mu_p}=\sum_a f^a_{\mu_1\ldots\mu_p}T^a$ where
$T^a$ are antihermitian generators of $U(N)$, this
necessitates taking the fields $f^a_{\mu_1\ldots\mu_p}$ to be
{\it complex}.
\item For a continuum gauge field we introduce lattice link 
fields $U_\mu(x)=e^{A_\mu(x)}=e^{A^a_\mu(x)T^a}$ with $A^a_\mu(x)$
complex together with 
its conjugate
$U^\dagger_\mu=e^{A^\dagger_\mu(x)}$. 
\item A covariant forward difference operator can be defined which acts on a field
$f_{\mu_1\ldots\mu_p}(x)$ as follows
\beq
D^+_\mu f_{\mu_1\ldots\mu_p}(x)=
U_\mu(x)f_{\mu_1\ldots\mu_p}(x+\mu)-
f_{\mu_1\ldots\mu_p}(x)U_\mu(x+e_{\mu_1\ldots\mu_p})\eeq
This operator acts like a lattice exterior derivative with
respect to gauge transformations in mapping a $p$-form lattice field
to a $(p+1)$-form lattice field.
\item Similarly we can define an adjoint operator $D^-_\mu$ 
whose action on some field
$f_{\mu_1\ldots\mu_p}$ is given by
\beq
D^-_\mu f_{\mu_1\ldots\mu_p}(x)=
f_{\mu_1\ldots\mu_p}(x)U^\dagger_\mu(x+e_{\mu_1\ldots\mu_p}-\mu)-
U^\dagger_\mu(x-\mu)f_{\mu_1\ldots\mu_p}(x-\mu)\eeq
It thus acts like the adjoint of the exterior derivative.
\item As for two dimensions all instances of $\partial_\mu$ in the
continuum action will be replaced by $D^+_\mu$ if the derivative
acts like $d$ (curl-like operation) and $D^-_\mu$ if the derivative acts 
like $d^\dagger$ (divergence-like operation).
\end{itemize}
This prescription preserves the twisted scalar supersymmetry and gauge
invariance and targets a complexified version of the 
${\cal N}=4$ super Yang-Mills model.
Again, to obtain ${\cal N}=4$ SYM we conjecture that the
Ward identities corresponding to
the scalar supersymmetry will be satisfied {\it without fine tuning} in the
continuum limit $\beta\to\infty$ for the truncated lattice model
in which the fields are taken real. 

\section{Conclusions}
In this paper we have presented a new way to 
put supersymmetric theories on the
lattice. The approach only works for theories with a number of supercharges
which is a integer multiple of $2^D$ if $D$ is the dimension of 
(Euclidean) spacetime. This includes the important cases of ${\cal N}=2$
and ${\cal N}=4$ super Yang-Mills theory in two and four dimensions
respectively.
The basic idea is to start from a {\it twisted} formulation of
the supersymmetric theory in which the fermions are represented
as grassmann valued antisymmetric tensor fields. The original supersymmetry
algebra is replaced by a twisted algebra which contains a scalar nilpotent
supercharge $Q$. Furthermore the action of the theory can then be written as
the $Q$-variation of some function.
It is straightforward to generalize the action of the scalar
supersymmetry to the lattice while maintaining its nilpotent
property. Invariance of the lattice theory under the scalar supersymmetry
can then be obtained easily. We have described the details of our
discretization prescription in detail. It requires the introduction
of particular covariant forward and backward difference operators and non standard
gauge transformation properties of the fields. The final theory is gauge
invariant, free of spectrum doubling and possesses an exact supersymmetry.
The price one pays for this is that the lattice fields are in general
complexified. We have argued that the $Q$-exactness of the action ensures
nevertheless that the theory truncated to the real line will satisfy the Ward
identities corresponding to the scalar supersymmetry without fine
tuning in the continuum limit. We have also presented preliminary
numerical results stemming from a dynamical fermion simulation of the
$D=2$ $U(2)$ model using
the new RHMC algorithm.

There are several directions for further work. 
The most obvious is the need for both perturbative and numerical checks
of both the exact and broken Ward identities
corresponding to the twisted supersymmetry. 
Exactly how much residual fine tuning would be
required in the four dimensional theory is currently unclear and could
be revealed by such an analysis.
It should also be possible to use these ideas to
construct lattice regularizations for the ${\cal N}=4,8$ theories
in two dimensions and the ${\cal N}=8$ theory in three dimensions.
This would allow contact to be made with predictions from 
string theory and supergravity theories in
lattice models which could be simulated relatively easily and with
high precision. 
The geometrical nature of the twisted theory lends one to speculate that it
should be possible to formulate these models on general simplicial
lattices. If this could be realized, then the resulting
models could be coupled to gravity in the usual way by
carrying out a sum over random lattices resulting in a lattice
regulated supergravity theory. It would also be very interesting to
understand the relationship, if any, between twisting and the orbifold constructions
of Kaplan at al. 

It appears that the twisted formulation of a theory with extended
supersymmetry affords a more useful starting point for
constructing lattice models over conventional approaches based
on spinors. However, 
the precise details of the discretization procedure are not determined
uniquely. While this work was in preparation a new paper by
D'Adda et al \cite{2dfullQ} was posted which claims to maintain
{\it all} the supercharges of the two dimensional twisted theory by
adopting a different mapping of the continuum twisted theory to the
lattice. This appears to be a very intriguing development.


\begin{thebibliography}{99}
\bibitem{wein} The Quantum Theory of Fields Vol. III Supersymmetry, S. Weinberg,
Cambridge University Press (2000).
\bibitem{prop} N. Seiberg and E. Witten, Nucl. Phys. B431 (1994) 484.
\bibitem{dual} E. Martinec and V. Sahakian, hep-th/9810224, L. Susskind,
hep-th/9805115, O. Aharony, J. Marsano, S. Minwalla and T. Wiseman,
hep-th/0406210, N. Itzhaki, J. Maldacena, J. Sonnenschein and S. Yankielowicz,
hep-th/9802042.
\bibitem{M} J.M. Maldecena, Adv. Theor. Math. Phys. 2 (1998) 231 [Int. J. Theor.
Phys 38 (1999) 1113].
\bibitem{general} P. Dondi and H. Nicolai, Nuovo Cimento 41A (1977)\\
S. Elitzur and A. Schwimmer, Nucl. Phys. B226 (1983) 109
J. Bartels and J. Bronzan, Phys. Rev. D28 (1983) 818.\\
T. Banks and P. Windey, Nucl. Phys. B198 (1982) 226. \\
I. Montvay, Nucl. Phys. B466 (1996) 259. \\
M Golterman and D. Petcher, Nucl. Phys. B319 (1989) 307.\\
W. Bietenholz, Mod. Phys. Lett A14 (1999) 51.
H. Aratyn and A.H. Zimerman, J. Phys. A 18 (1985) L487.
(1985) 225\\
Y. Kikukawa and Y. Nakayama, Phys. Rev. D66 (2002) 094508.\\
K. Fujikawa, Phys. Rev. D66 (2002) 074510.\\
J. Nishimura, Phys. Lett. B406 (1997) 215.\\
S. Catterall and S. Karamov, Phys. Rev. D 68 (2003) 014503.\\
W. Bietenholtz, Mod. Phys. Lett. A14 (1999) 51.\\
M. Beccaria, M. Campostrini, G.De Angelis and A. Feo, Phys. Rev. D70 (2004)\\
M. Beccaria, M. Campostrini, A. Feo, Phys. Rev. D69 (2004) 095010.\\
H. Aratyn, M. Goto and A.H. Zimerman, Nuovo Cimento A88
\bibitem{feo} A. Feo, Supersymmetry on the Lattice, hep-lat/0210015\\
\bibitem{kaplan} D. B. Kaplan, Recent Developments in Lattice Supersymmetry,
hep-lat/0309099
\bibitem{feowz} M. Bonini and A. Feo, JHEP 0409 (2004) 011.
\bibitem{moto} $N=1$ super Yang-Mills on a (3+1) dimensional 
transverse lattice, M. Harada, S. Pinsky, hep-lat/0411024, M. Harada, S. Pinsky,
Phys. Rev. D70 (2004) 087701, M. Harada, S. Pinsky, Phys. Lett. B567 (2003) 277
\bibitem{kap}
D.B. Kaplan, E. Katz and M. Unsal, JHEP 0305 (2003) 037, A.G. Cohen, D.B. Kaplan, E. Katz, M. Unsal, 
JHEP 0308 (2003) 024, A.G. Cohen, D.B. Kaplan, E. Katz, M. Unsal, JHEP 0312
(2003) 031
\bibitem{kap4} D. B. Kaplan and M. Unsal, hep-lat/0503039.
\bibitem{unsal} M. Unsal, hep-lat/0504016.
\bibitem{others}
J. Nishimura, S. Rey and F. Sugino, JHEP 0302 (2003) 032\\
J. Giedt, E. Poppitz and M. Rozali, JHEP 0303 (2003) 035\\
J. Giedt, Nucl. Phys. B668 (2003) 138\\
J. Giedt, Nucl. Phys. B674 (2003) 259.
\bibitem{top} S. Catterall, JHEP 0305 (2003) 038.
\bibitem{joel} J. Giedt, E. Poppitz, hep-lat/0407135.
\bibitem{kaw2dtwist} A. D'Adda, I. Kamamori, N. Kawamoto, K. Nagata,
hep-lat/0406029.
\bibitem{qm} S. Catterall and E. Gregory, Phys. Lett. B487 (2000) 349.
\bibitem{qm_joel} Giedt, Koniuk, Poppitz and Yavin, JHEP 0412 (2004) 033.
\bibitem{wz2} S. Catterall and S. Karamov, Phys. Rev. D65 (2002) 094501.
\bibitem{wz2_joel} Giedt, hep-lat/0507016.
\bibitem{sigma} S. Catterall and S. Ghadab, JHEP 0405 (2004) 044.
\bibitem{sugino1} F. Sugino, JHEP 0401 (2004) 015.
\bibitem{sugino2} F. Sugino, JHEP 0403 (2004) 067.
\bibitem{tqft} E. Witten, Comm. Math. Phys. 117 (1988) 353.
\bibitem{tak} N. Kawamoto , T. Tsukioka, Phys. Rev. D61 (105009) 2000.
\bibitem{kato} J. Kato, N. Kawamoto and Y. Uchida, Int. J. Mod. Phys. A19 (2004)
2149.
\bibitem{lat04} S. Catterall, Nucl. Phys. B Proceedings Suppl. 140 (2005) 751
\bibitem{2dsuper} S. Catterall, JHEP 0411 (2004) 006.
\bibitem{toprev} D. Birmingham, M. Blau, M. Rakowski and G. Thompson,
Phys. Rep. 209, numbers 4 and 5 (1991) 129.=
\bibitem{banks} T. Banks, Y. Dothan and D. Horn, Phys. Lett. B117 (1982) 413.
\bibitem{4dsuper} S. Catterall, JHEP 06 027 (2005).
\bibitem{kaw4dtwist}
D'Adda, Kanamori, Kawamoto and Nagata, Nucl. Phys. B707 (2005) 100.
\bibitem{schwim} S. Elitzur, E. Rabinovici and A. Schwimmer, Phys. Lett B119 (1982) 165.
\bibitem{adjoint} H. Aratyn, M. Goto and A.H. Zimerman, Nuovo Cimento A84,
(1984) 255
\bibitem{becher} P. Becher, Phys. Lett. B104 (1981) 221\\
P. Becher and H. Joos, Z. Phys. C 15 (1982) 343.
\bibitem{rabin} J. Rabin, Nucl. Phys. B201 (1982) 315.
\bibitem{rhmc} M. Clark and A. Kennedy, hep-lat/0409133.
\bibitem{hmc} S. Duane, A. Kennedy, B. Pendleton and D. Roweth, Phys. Lett.
B195B (1987) 216. 
\bibitem{multi} B. Jegerlehner, Krylov solvers for shifted linear
systems hep-lat/9612014
\bibitem{soon} Numerical study of twisted ${\cal N}=2$ 
super Yang-Mills, S. Catterall, in preparation. 
\bibitem{marcus} N. Marcus, Nucl. Phys. B452 (1995) 331.
\bibitem{4dspinor} E. D'Hoker and D. Freedman, hep-th/0201253.
\bibitem{2dfullQ} D'Adda, Kanamori, Kawamoto and Nagata, hep-lat/0507029.
\end{thebibliography}
\end{document}